\documentclass[authoryear,final,5p,times,twocolumn]{elsarticle}
\usepackage{color}
\usepackage{graphicx}
\usepackage{amsmath,amssymb, amsthm}
\newcommand{\be}{\begin{equation}} \newcommand{\ee}{\end{equation}}
\newcommand{\ba}{\begin{align*}} \newcommand{\ea}{\end{align*}}
\newcommand{\bea}{\begin{eqnarray}} \newcommand{\eea}{\end{eqnarray}}
\newcommand{\beaa}{\begin{eqnarray*}} \newcommand{\eeaa}{\end{eqnarray*}}
\journal{Theoretical Population Biology}
\begin{document}
\begin{frontmatter}
\title{Evolutionary branching in a stochastic population model with discrete mutational steps}
\author[adr1]{S.~Sagitov}
\author[adr2]{B.~Mehlig\corref{cor1}}
\ead{Bernhard.Mehlig@physics.gu.se}
\author[adr1]{P.~Jagers}
\author[adr3]{V.~Vatutin}
\address[adr1]{Mathematical Sciences, Chalmers and Gothenburg University,  SE-41296 Gothenburg, Sweden.}
\address[adr2]{Department of Physics, University of Gothenburg, SE-41296 Gothenburg, Sweden.}
\address[adr3]{Department of Discrete Mathematics, Steklov Mathematical Institute, Moscow, Russia}
\cortext[cor1]{Corresponding author}
\begin{abstract}
Evolutionary branching is analysed in a stochastic, individual-based
population model under mutation and selection.
In such models, the common assumption is that individual reproduction and life
career are characterised by values of a trait, and also by population
sizes,  and that mutations lead to
small changes $\epsilon$ in trait value. Then, traditionally,
the evolutionary dynamics is studied in the limit $\epsilon\rightarrow 0$.
In the present approach, small but non-negligible mutational steps are
considered. By means of theoretical analysis 
in the limit of infinitely large populations,
as well as computer simulations,
we demonstrate how discrete mutational steps affect
the patterns of evolutionary branching. We also argue that the average
time to the first branching depends in a sensitive way on both
mutational step size and population size.
\end{abstract}
\end{frontmatter}
\section{Introduction}\label{sec:in}
The development of populations subject to
frequency- or density-dependent selection is an important and central but not easily
analysed topic in theoretical biology. Many different models and
approximation schemes have been developed in order to understand how
trait values change and how the intriguing possibility of \lq evolutionary
branching' may arise, where a population of individuals with a well-defined
trait value or genotype (monomorphic, in other words) experiences a
split leading to two or more coexisting trait values or genotypes in the
population. This process is important because it gives rise to increased biodiversity.

A variety of population-genetic and game-theoretic methods, as well as 
the body of ideas and approximation schemes known as  {\em adaptive dynamics},
cf. \cite{M96,G98,Odo,WG,MAD}, have been employed in order 
to study evolutionary branching.
Several simplifications are commonly resorted to.  
First, complications due to mating and recombination in 
diploid populations are disregarded by ascribing to each individual 
a single trait value $x$ that is transferred by clonal reproduction 
unless mutation occurs \citep[see however the recent interesting approach 
to adaptive dynamics with Mendelian inheritance by][]{MeMe}.
Selection is taken to act through a trait- and density-dependent
fitness function, identified with mean reproduction. Second, it is
assumed that fitness is a smooth function of the trait value $x$ and the population
density. The third postulate is that the mutation rate $\mu$ is so
small that only few trait values are represented in the population
at any one time. Mutations thus occur so rarely that we can talk of a
separation of time scales between long-term evolutionary dynamics (the
sequence of mutations) and short-term population
dynamics, defined by birth and death events, an evolutionary versus an
ecological time scale. Fourth, it is assumed that mutations lead 
only to small changes in the trait value (that mutational step sizes
$\epsilon$ are small). Fifth, 
the population size $N$ is taken to be large, and effects of random genetic drift
are neglected (apart from incorporating the fixation probability of an advantageous mutation).

In the adaptive-dynamics literature a further step is taken by letting $N\rightarrow \infty$,
 $\mu\rightarrow 0$, and $\epsilon \rightarrow 0$. Then evolution of a
 monomorphic population to the first branching of traits is analytically
described by the so-called \lq canonical equation' \citep{DL}, or more generally
by deterministic evolution on the fitness landscape \citep{Wax2,Lande,Iwasa}. 
How these deterministic adaptive dynamics may arise from a strict
stochastic population model has been described in detail in the
elegant treatment of \citet{Ch} and later papers. 

As a monomorphic (single-trait) population approaches a local fitness
maximum where evolutionary branching may occur in the sense that
populations with two different trait values may coexist from there
onwards, the canonical equation fails to describe the dynamics. It 
becomes necessary to ask how stochasticity in a finite population
affects the possibility of evolutionary branching. Further questions
are: what is the shape of the bifurcation diagram of trait values in
the wake of a branching? How do the values of $\mu$ and $\epsilon$
influence the evolutionary dynamics in a finite population? How long
does it  take (in the mean) for evolutionary branching to occur? 

We address these matters analytically and by computer simulations of a
stochastic, individual-based  model for the evolution of trait values
in a finite population subject to density-dependent fitness function symmetric with respect to its maximal value at the trait $x=0$.  In the limit of $N\rightarrow
\infty$, $\mu\rightarrow 0$, and $\epsilon$ small, we show
that the evolution of a monomorphic population towards the 
fitness optimum at $x=0$ is governed by a canonical equation. For our model, standard stability analysis in the limit of $\epsilon \rightarrow 0$ would predict evolutionary branching of a monomorphic population at 
$x=0$. 
We demonstrate that at small but fixed mutational step sizes $\epsilon$,  
evolutionary branching typically occurs at non-zero values of $x$.
In the limit of infinite population size and vanishing mutation rate 
we compute the critical value of $x$ where the first branching is most
likely to occur, as a function of $\epsilon$.
In addition we show that the evolutionary dynamics in this limit
gives rise to a non-symmetric bifurcation diagram after the first branching
(for a symmetric model such as ours, the standard theory predicts a symmetric bifurcation diagram
in the limit  $\epsilon \rightarrow 0$).
Moreover, we state geometrical conditions determining the trait values
where the second evolutionary branching event is most likely to occur.
 
Finding the average time until evolutionary branching occurs in our model requires a more refined analysis. 
It is necessary
to specify how  $N\rightarrow \infty$ and $\epsilon\rightarrow 0$.
This determines the stochastic population dynamics in the critical
stage preceding the first branching. That stage must last  long enough
to render branching possible. We discuss results of computer
simulations relating to this question briefly, but a precise
mathematical description of the problem is left for future work.

The remainder of this paper is organised as follows.
In Section \ref{sec:model} we introduce the stochastic, individual-based model 
used in  simulations. Section \ref{sec:AD} summarises standard results
obtained in the limit $N\rightarrow \infty$, $\mu\rightarrow 0$, and $\epsilon\rightarrow 0$. These  are contrasted with numerical experiments at relevant
values of $N$, $\mu$, and $\epsilon$ in Section \ref{sec:sub4}.  
Discrepancies between the $\epsilon\rightarrow 0$
predictions and the results of numerical experiments
are discussed. Our results (and the calculations supporting them) addressing the issues raised in Section \ref{sec:sub4} are
summarised in Section \ref{sec:results}.  Section \ref{sec:conc}
contains our conclusions. Three appendices summarise details of our 
calculations.

\section{Model}

\begin{figure}[tbp]
\includegraphics[width=80mm]{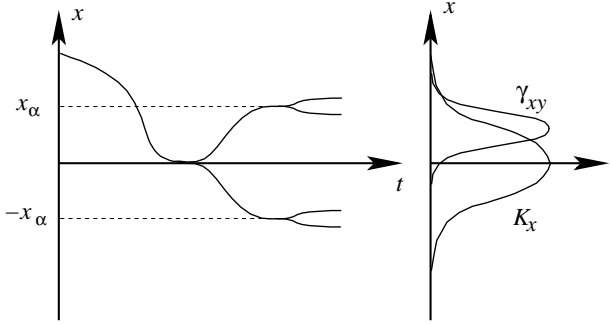}
\caption{
A schematic view of the deterministic model. The left panel depicts the first three evolutionary
branching events. The right panel illustrates the main ingredients of the fitness function given by Eq. (\ref{eq:K}). The function $\gamma_{xy}$ is plotted for a particular value of $y=x_0$ which in the case shown is positive.}
\label{fdet}
\end{figure}

\label{sec:model}
The stochastic, individual-based  model
used in the numerical experiments discussed below describes a
finite population of individuals competing for resources.
Each individual is assigned a trait value $x$. Individuals with the same
trait value (belonging to the same ecological \lq niche') compete for a limited
amount of resources. The scarcity of resources is parametrised
by a trait-dependent carrying capacity $K_x$. There is an \lq optimal'
trait value corresponding to the value of $x$ where $K_x$ assumes its
global maximum. Individuals belonging to different niches
(corresponding to different trait values, $x$ and $y$, say) 
may also interact. The strength of interaction depends
upon the difference $|x-y|$ in trait values.

Mutations give rise to changes in the trait value, 
$x\rightarrow x\pm \epsilon$. 
Here $\epsilon>0$ is a fixed mutational step size and the set of
possible traits occurring in the population is
$x=0,\pm\epsilon,\pm2\epsilon,\ldots$.
Individual fitness is measured by the mean offspring number, 
which is determined by the trait values and the intensity of competition. The
latter, in its turn,  is controlled by the scarcity of resources 
and the number of individuals. In other words, the model describes the
evolution of a trait value, or of trait values, in a population subject
to density-dependent selection.  The population would be expected 
to evolve towards the optimal trait value. Intense competition between 
individuals with trait values close to the optimal one may however
give rise to sub-populations with well-defined but distinct trait
values. This phenomenon is referred to as  {\em evolutionary
  branching}, and is the subject of this paper.

The number of individuals with a given trait value $x$ 
is denoted by $Z_x$. It is assumed that each individual 
has either none or two offspring  \citep{JBD}. The offspring of individuals in one
generation constitute the next generation. We denote the fitness function by 
$M_x(Z_y,y\in D)$ which is
the mean offspring number for an individual with
trait value $x$, and where $D$ is the
set of trait values represented in the population. This fitness is a function of the co-existing
sub-population sizes ($Z_y, y \in D)$. We take it to be of the form:
\begin{equation}
M_x(Z_y,y\in D)=\frac{2}{1+({NK_x})^{-1}\sum_{y\in D}\gamma_{xy}Z_{y}}\,.
\label{Mz}
\end{equation}
Here the carrying capacity of the subpopulation with trait value $x$ 
is $N K_x$, where $N$ is a population-size parameter to be thought of
as large. Competition between subpopulations corresponding to
different trait values $x$ and $y$ is regulated by the function $\gamma_{xy}$.

The offspring usually inherit the parental trait $x$,  
but with a small probability $\mu$ they mutate (independently)
either to $x-\epsilon$ or to $x+\epsilon$, with equal probabilities. 
Since time is measured in generations, the model is 
thus discrete both in time and trait space. It can be viewed as the simplest possible structure \citep[`the bare bones', ][]{JBD} behind adaptive dynamics, and as such is an interesting mathematical object in its own right. Since all individuals have the same life span, one, it displays an unyielding age dependence which can be seen as a diametrical counterpart and natural complement to the equally artificial non-aging individuals of birth-and-death processes, 
underlying the sequence of papers by \citet{CM} and others, and also classical deterministic approaches. For more realistically age-structured models cf. \cite{Tran} as well as  \cite{Jakle}.

The form \eqref{Mz} is certainly ad hoc, and should be viewed as an
explorative illustration. But it is not without historical precedence \citep{CL}. 
As there,  we take $K_x$ and $\gamma_{xy}$ to be Gaussian functions. 
Indeed, this is a choice often encountered in the adaptive-dynamics literature,
\begin{equation}
\label{eq:K}
K_x = {\rm e}^{-x^2}\quad\mbox{and}\quad \gamma_{xy} = {\rm e}^{-(x-y)^2(1+\alpha)}\,.
\end{equation}
The parameter $\alpha$ regulates the competition between
subpopulations with different traits. A larger value of $\alpha$
implies  weaker competition. We take $\alpha > 0$. This ensures that competition between
subpopulations is weak enough to allow for evolutionary branchings. It
is convenient to replace  population sizes $Z_x$ by  densities
$f_x={Z_x}/{(NK_x)}$, so that the fitness function \eqref{Mz} takes
the form
\begin{equation}
M_x(f_y,y\in D)=\frac{2}{1+\sum_{y\in D}R_{xy}f_{y}}\,.
\label{Mx}
\end{equation}
Here
\begin{equation}\label{MR}
R_{xy}=\gamma_{xy}K_y/K_x=e^{-(x-y)(\alpha x-(2+\alpha)y)}
\end{equation}
is the fitness kernel which combines the effects of the interaction $\gamma_{xy}$ and the carrying capacity $K_x$. 

Eqs.~(\ref{Mx}) and \eqref{MR} allow for  explicit formulae for the equilibrium
densities $(f_y^D,y\in D)$ at which subpopulations may coexist in the absence of mutations. Note however that even this coexistence is temporary, 
albeit potentially of long duration \citep{Jakle}. We shall still refer to them as equilibrium densities. They are defined by the condition $M_{x}^D=1$, where
\be
M_{z}^D\equiv\frac{2}{1+\sum_{y\in D}R_{zy}f^D_{y}} \,.
\label{MD}
\ee
The equation $M_{x}^D=1$ describes the critical regime of reproduction. It
corresponds to the condition
\be
\sum_{y\in D}R_{xy}f^D_{y}=1,\ x\in D \,.
\label{MD1}
\ee
In the monomorphic case $D=\{x\}$ we find that
$M_x(f_x)=2/(1+f_x)$ and in the absence of mutations, the density $f_x$ stabilises at 
the value $f^{\{x\}}_x=1$. 
Substituting this result into Eq.~(\ref{MD}) we obtain 
\begin{equation}\label{fd}
M_z^{\{x\}}={2\over 1+e^{-(z-x)(\alpha z-(2+\alpha)x)}}\,.
\end{equation}

Importantly, the function $M_z^D$ given by Eq.~(\ref{MD}) plays a crucial role in the invasion analysis. Suppose an individual of type $z\notin D$ appears 
in a stable population with the set of traits $D$ as a  mutant of $x\in D$. 
Its fitness has the form
\begin{align*}
M_{z}(f_y^D,y\in D;f^*_z)&=\frac{2}{1+\sum_{y\in D}R_{zy}f_{y}^D+f_z^*}\approx M_z^D\,
\end{align*}
because $R_{zz}=1$ and 
the small initial frequency of the mutant $f_z^*$ can be neglected.  If $M_z^D>1$, there is a chance for the new trait to establish itself in the set of coexisting traits. If $M_z^D\le1$, 
the mutation can not establish itself and is wiped out by genetic drift.

In summary, we have chosen an extremely simple model which however
retains so much of the structure of evolutionary dynamics that the
questions raised in the introduction can still be formulated. 
It is described by only 
four parameters, the population-size parameter
$N$, the mutation rate $\mu$, the mutational step size $\epsilon$, and
the interaction parameter $\alpha$.

\section{Standard predictions in the limit $\epsilon \rightarrow 0$}              
\label{sec:AD}
In spite of its simplicity, the  model introduced in the previous Section
 is not easily analysed. 
 As pointed out, the standard adaptive-dynamics approach \citep{G98} studies continuous-time 
population models in the limit $N\to\infty$  and $\mu\to 0$, and then $\epsilon\to 0$. 
In this Section we state the corresponding results for our discrete-time model.
  
If $N$ can be thought of as infinite, law-of-large-number effects
replace random processes by their expectations, and if both
$\mu\rightarrow 0$ and $N\rightarrow \infty$,  then only a small number of
trait values are represented in the population  at
any specific time: the basic cases studied below are those of monomorphic
populations, $D=\{x\}$,
and dimorphic ones, $D=\{x_1,x_2\}$.

In this limit, a time-scale separation appears
between the evolutionary dynamics (referring to long-term changes
of the trait value) and the ecological time scale of population
dynamics. 
Letting $\epsilon\rightarrow 0$ makes it possible to describe the dynamics
of the trait value $x$ by local stability analysis, central to
adaptive dynamics.

Fig.~\ref{fdet} illustrates the standard predictions obtained in
these limits.  First, starting from an initially positive trait value, $x(t)$ 
is expected to evolve deterministically
towards the local optimum (at $x=0$ in our model). 
The path $x(t)$ is described by a differential equation referred to as 
a `canonical equation'
\citep{DL,CFB} and related to similar equations studied in population genetics \citep{Lande,Iwasa}.
Second, the first evolutionary branching occurs at $x=0$. Third, after the first
branching, the pair of traits evolves deterministically in a symmetric fashion, 
and the second and third branchings occur simultaneously.

\subsection{Deterministic evolution towards the first branching}
\label{sec:sub1}
Assume that the population starts from a positive 
initial trait value $x_0=j_0\epsilon$ which
is large compared to the small value of the mutational step size
$\epsilon$: $j_0 \gg 1$. Classical adaptive dynamics then
predicts that the initial trait value is consecutively replaced by
smaller values of $x$, driving $x(t)$ towards $x=0$. 
The resulting function of time $x(t)$ is described by a differential
equation for $x(t)$, known as the canonical equation (of trait
evolution), the standard reference being \cite{DL}. In the notation of  \citet[see Eq.~(7a) ]{CFB} the classical canonical equation of trait
evolution has the form
\begin{equation}
\label{eq:cest}
{{\rm d}x\over {\rm d}t}= \mu(x) {\sigma_0^2\over2}n(x)\partial_1f(x,x)\,.
\end{equation}
In this expression, $\mu(x)$ is the mutation rate for the trait value $x$,
$\sigma_0^2$ is the mutational variance, $n(x)$ is the equilibrium population size, and  $f(z,x)$ is the fitness function.  
Recall that 
within a model formulated in terms of stochastic birth and death processes,
the fitness function $f(z,x)$ is the difference between the birth and death rates for rare mutants 
with trait $z$ in an $x$-population at its quasi-stable size. 

In our case, the mutation rate $\mu(x)=\mu$ is independent of the trait value $x$,
$\sigma_0^2$ is simply $\epsilon^2$, and $n(x)=Ne^{-x^2}$.
The fitness function $f(z,x)$ corresponding to our discrete-time model should be calculated as
\begin{equation}\label{fc}
f(z,x)=\log M_z^{\{x\}},
\end{equation}
where  $M_z^{\{x\}}$ is given by Eq.~(\ref{fd}). This formula relies 
on a well-known property of
a linear birth-death process stemming from a single individual: if $f$ is the difference between the birth and death rates per individual, the population size at time $t$ has mean $e^{tf}$. Eq.~(\ref{fc}) stipulates that the expected population size $e^{f(z,x)}$ after one unit of  continuous time is equal to the expected size $M_z^{\{x\}}$ of the first generation in our discrete-time model. 

From Eqs.~(\ref{fc}) and (\ref{fd}) we conclude that the fitness gradient  is 
given by
$$\partial_1f(x,x)=-x\,,$$ 
since $M_x^{\{x\}}=1$.
In our case we would thus
expect the canonical equation to take the form
\begin{equation}\label{CEq}
{{\rm d}x\over {\rm d}t}= - \mu{\epsilon^2\over2} N{\rm e}^{-x^2}x\,.
\end{equation} 
However, the correct canonical
equation for our model  Eq.~(\ref{CEq2}) derived in Section \ref{sec:standard} reveals that the trait substitution process goes twice as fast in the 
discrete-time model compared to the  continuous-time model. 
This is reminiscent of the well-known phenomenon of genetic drift running twice as fast in the Moran model than in the Wright-Fisher model of the same size, see for example \cite{Wa}, though the latter phenomenon has a different explanation.

\subsection{Location of the first evolutionary branching}
\label{sec:sub2}
The canonical equation  
suggests that the substitution process 
of the trait value slows down as the point $x=0$ is approached. 
In the limit of $N\to\infty$, $\mu\to 0$, and $\epsilon\to 0$, standard adaptive-dynamics 
analysis predicts the first evolutionary branching to occur at $x=0$, 
giving rise to two branches, $x_1 <0 $ and $x_2 > 0$. 

\subsection{Second and third evolutionary branchings}
\label{sec:sub3}
Moreover, in the very same limit, the deterministic evolution after the 
first branching is predicted to be symmetric. In other words,
the two trait values $x_1$ and $x_2$ after the first branching are expected to 
evolve as  $x_1(t) = -x_2 (t)$. 
The second and third evolutionary branchings
are expected to occur simultaneously when $x_2(t)$ reaches
a critical value, which in our case 
can be computed explicitly as  
\begin{equation}
\label{xa}
 x_\alpha={1\over2}\sqrt{{\log(2\alpha+1)\over 1+\alpha}}\,.
\end{equation}
As a function of $\alpha$ the value of $x_\alpha$ reaches its maximum $0.3731\dots$ at $\alpha=1.2955\dots$.

\section{Computer simulations}
\begin{figure}[t]
\includegraphics[width=80mm]{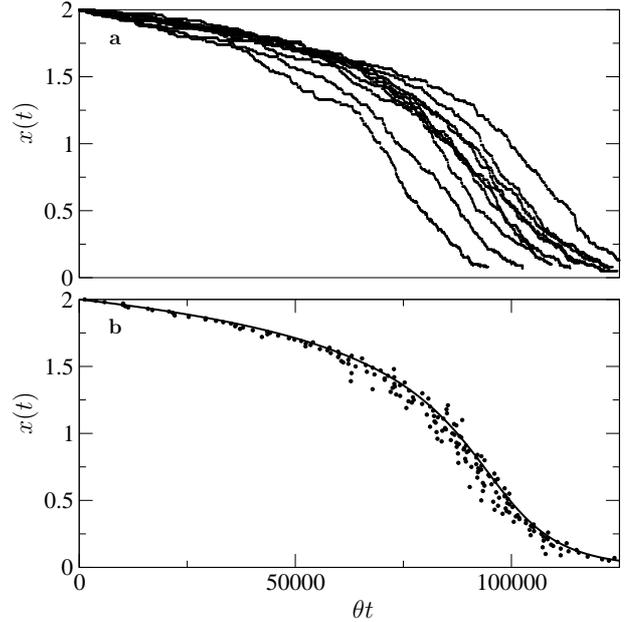}
\caption{\label{fig:ceq} 
{\bf a} Shows ten realisations of the evolution of $x(t)$
from an initially large value $x_0=2$ for a mutational step size of
$\epsilon = 10^{-2}$ (additional parameters:  $N=10^6$, $\mu=10^{-8}$,
$\alpha=9$, $\theta = \mu N = 10^{-2})$.
{\bf b}  Shows the trait value $x$ as  a function of
the average time of staying at $x$, 
conditional on that no branching has occurred (dots).
Also shown is an approximate solution of
\eqref{CEq} (solid line).}
\end{figure}

\begin{figure}[t]
\includegraphics[width=80mm]{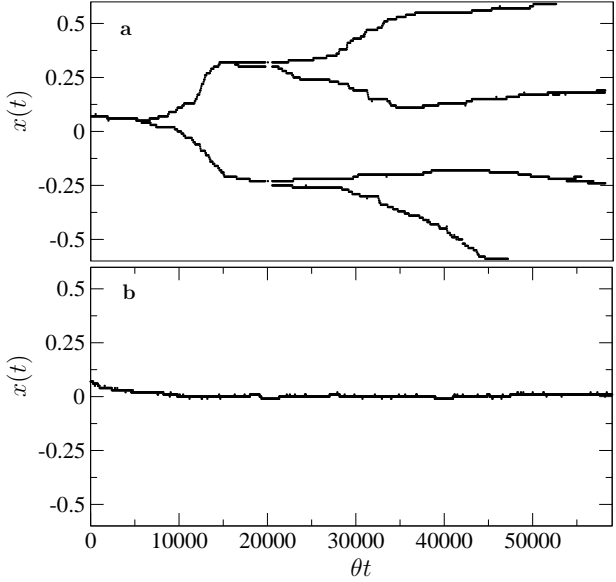}
\caption{Results of computer simulations of trait evolution in the the stochastic individual-based
model described in Section~\ref{sec:model}. Parameters:
$\alpha = 9$, $\epsilon   = 0.01$, and {\bf a} $N=5\times 10^6$,
$\mu =2\times 10^{-10}$. 
}
\label{Fsim}
\end{figure}

\begin{figure}[t]
\includegraphics[width=60mm]{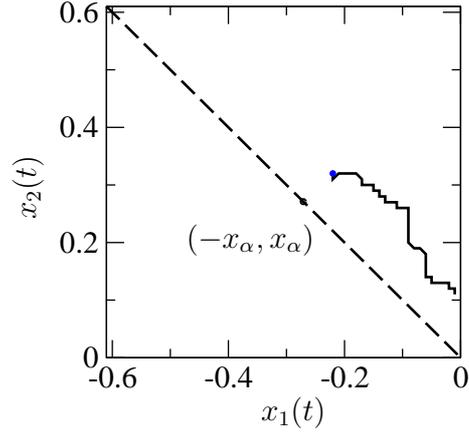}
\caption{Shows the trajectory of Fig.~\ref{Fsim}{\bf a}
after the first branching in the $x_1$-$x_2$-plane.
Only points corresponding to $x_1 < 0$ are shown.
Parameters the same as Fig.~\ref{Fsim}{\bf a}.  }
\label{Fsim2}
\end{figure}

\label{sec:sub4}
Figs.~\ref{fig:ceq}, \ref{Fsim} and \ref{Fsim2} summarise results of direct numerical simulations of the model described in Section~\ref{sec:model} 
for large values of $N$, 
small values of $\mu$, and for a small value of $\epsilon$ (equal to $10^{-2}$). 
In the algorithm, binomial distributions of numbers of mutants have been approximated by the 
appropriate Poisson distributions.

We now compare simulation results with the predictions from Section \ref{sec:AD}.
Fig.~\ref{fig:ceq}{\bf a} shows ten realisations of the evolution of $x(t)$
for the mutational step size $\epsilon = 10^{-2}$ from an initially large value $x_0=2$. (The additional parameters are given in the figure caption). We see that $x(t)$ decreases
towards the first branching (which in all cases occurs for positive values of $x$ and not at $x=0$).
Fig.~\ref{fig:ceq}{\bf b} shows a plot of the trait value $x$ versus
the average time of staying at $x$, conditional on  no branching
having occurred. Also shown is an approximate solution of the correct
canonical equation derived in Section \ref{sec:standard} as
Eq.~\eqref{CEq2}. This is expressed in terms of the expected time for  $x(t)$ to reach level $x$:
\begin{equation}
\label{eq:tav}
t(x) = \sum_{y=x-\epsilon}^{x_0} \lambda_y^{-1}\,,
\end{equation} 
where $\lambda_x = \mu N \epsilon x {\rm e}^{-x^2}$ (see below) is the rate of 
the asymptotically exponential holding time  at level $x$.
For large values of $x$ we observe good agreement between (\ref{eq:tav})
and the average of ten independent realisations of the substitution process. 
But as $x=0$ is approached, the numerical average falls below the predicted average. The neighbourhood of the first branching event is not described by the canonical equation.

Fig.~\ref{Fsim}
shows results of two computer simulations starting in the vicinity
of $x=0$. In Fig.~\ref{Fsim}{\bf a} the first branching does not occur
at $x=0$ but at $x=0.05$. Further, evolution after the first branching
is not symmetrical. This can also be seen from Fig.~\ref{Fsim2} which
displays the data of Fig.~\ref{Fsim}{\bf a} in the $x_1$-$x_2$ plane
(the standard adaptive-dynamics approach predicts, as pointed out above,
that the pair ($x_1(t),x_2(t)$) moves on the diagonal shown in Fig.~\ref{Fsim2} as a dashed line).
Note also that the second and third branchings do not occur simultaneously, and
not symmetrically in the simulations.

Fig.~\ref{Fsim}{\bf b} shows results of a direct numerical simulation
with the same overall substitution rate $\theta = \mu N$ as Fig.~\ref{Fsim}{\bf a},
but for a smaller value of $N$ and a larger value of $\mu$. In this example,
branching did not occur during the simulation time $\theta t < 10^5$. By contrast,
the trait value is seen to diffuse around $x=0$. 

In summary we find that the standard adaptive-dynamics approach describes the initial time evolution
of the trait value $x(t)$ well, provided $x(t)$ is large enough. In the vicinity of
and after the first branching, however, the observed patterns of evolutionary dynamics 
(Figs.~\ref{Fsim} and \ref{Fsim2})
differ from the standard predictions.
In the following we show that the asymmetric patterns 
observed in Figs.~\ref{Fsim}{\bf a} and \ref{Fsim2} are due to the positive mutational
step size $\epsilon$. In the limit $N\rightarrow \infty$ and $\mu\rightarrow 0$
we were able to characterise the branching patterns at small but fixed values of $\epsilon$. 
To determine how long it takes on average until evolutionary
branching occurs in a finite population requires a more refined analysis.
We return to this in Section \ref{sec:3.5}.

\section{Results}
\label{sec:results}

\subsection{Deterministic evolution towards the first branching}
\label{sec:standard}

In the monomorphic case ($D=\{x\}$) Eq.~\eqref{fd} implies 
\be\label{1eps}
M_{z}^{\{x\}}-1\approx x(x-z)+\left({\alpha/2} -x^2\right)(z-x)^2\,,
\ee
for $z\approx x$.  For $x>0$ it follows that
\begin{equation}
M_{x-\epsilon}^{\{x\}}>1, \, M_{x+\epsilon}^{\{x\}}<1, \, M_{x}^{\{x-\epsilon\}}<1, \quad\mbox{and}\quad M_{x}^{\{x+\epsilon\}}>1\,.
\end{equation}
These inequalities show
 that the expected trend of adaptive evolution in the monomorphic phase is simple: the initial positive trait value is 
consecutively replaced by smaller values driving $x(t)$ towards zero.  The dynamics of this trait substitution process is described by the canonical equation whose correct form for our model is derived next.

Observe that the expected number of mutations $x\rightarrow x-\epsilon$ in the $x$-population, assumed to be at its
quasi-stable size, $Z_x=NK_x=Ne^{-x^2}$, is $(\mu/2) Ne^{-x^2}$ per generation, where $\mu$ is divided by 2 since we discard the  $x\rightarrow x+\epsilon$-mutations. Thus the expected waiting time between two consecutive replacements is
\begin{equation}
\label{eq:lambdax}
\lambda_x\approx{\mu\over2}Ne^{-x^2}\, P_{x-\epsilon}(x)\,.
\end{equation}
Here $P_z(x)$ stands for the
probability that a population from one $z$-individual survives and
takes over from the monomorphic $x$-population. 
To find the  probability $P_z(x)$ of the mutant $z=x-\epsilon$ invading
and replacing a resident population with trait $x$ we can approximate
the population dynamics of the invader by a binary branching process starting from a
single individual. The survival probability of this branching process is 
\begin{equation}
P_z(x)=2{ M_z^{\{x\}}-1\over M_z^{\{x\}}}\,,\label{spr}
\end{equation}
see for example Eq.~(5.66) in  \cite{HJV}. Eq.~\eqref{1eps} implies
\begin{equation}
M_{x-\epsilon}^{\{x\}}\approx1+\epsilon x\,,\label{exp}
\end{equation}
and we obtain
$P_{x-\epsilon}(x)\approx2\epsilon x$, at least for $x$ far from
zero. (The corresponding classical result for the Wright-Fisher model 
is the following. Suppose that an advantageous allele is introduced
in the population with its mean offspring number being
larger compared to the wild type by a factor  $(1+s)$.
Then the fixation probability of this allele is approximately
given by $P \sim 2s$ for small values of $s$.) 
 
Substituting $P_{x-\epsilon}(x)\approx2\epsilon x$ into Eq.~(\ref{eq:lambdax}) 
results in $\lambda_x\approx\mu\epsilon Ne^{-x^2}x$. Now, since the evolution of the trait value $x(t)$ as a function of time is approximately given by
\begin{equation}
-{{\rm d}t\over {\rm d}x}
\approx{t(x-\epsilon)-t(x)\over \epsilon}\approx{1\over \epsilon \lambda_x}\,,
\end{equation}
we conclude that the canonical equation takes the form
\begin{equation}
\label{CEq2}
{{\rm d}x\over {\rm d}t}= - \mu\epsilon^2 N{\rm e}^{-x^2}x\,
\end{equation} 
which is similar (but not identical) to Eq.~(\ref{CEq}).  
Indeed, Eqs.~(\ref{CEq2}) and (\ref{CEq}) differ by a factor of
$1\over 2$. 

This factor can be explained as follows.
The standard derivation of Eq.~(\ref{eq:cest}) \citep{DL,CFB} refers to birth-and-death 
processes in continuous time (or corresponding deterministic
formulations), whereas our model considers binary splitting in
discrete, non-overlapping generations.  
Our form of the equation appears explicitly for Poisson reproduction in discrete time in \cite{Col}. 
The fact that different effective population sizes can appear in the canonical equation was first noted 
by \cite{DM}.

Note
that the speed of the trait-substitution process given by
Eq.~(\ref{eq:cest}) would not change
if we were to take different values of the birth and death rates, as long as their difference $f(z,x)$ remains the same. We argue that the proper choice of the birth rate in the monomorphic stable population is 
unity, matching the birth rate in our discrete-time model: 
one birth per generation per individual, $M_x^{\{x\}}=1$. Then, in accordance with 
Eq.~(\ref{fc}), the corresponding survival probability of a single mutant is given by $f(x-\epsilon,x)=\log M_{x-\epsilon}^{\{x\}}\approx \epsilon x$. Comparing this with its discrete-time counterpart 
$P_{x-\epsilon}(x)\approx2\epsilon x$, we conclude that the survival probability in the discrete model is two times larger. 

An important assumption behind Eq.~(\ref{CEq2}) is that successful mutations 
should not come too soon after each other in order to make sure
that selective sweeps do not overlap (\lq clonal interference'), see \cite{Krug}.
As a very rough approximation, we estimate the average fixation time
from the equation  
\begin{equation}
\Big(M_{x-\epsilon}^{\{x\}}\Big)^{T_{\rm fix}}\approx N.
\label{m2}
\end{equation}
This equation simply suggests to neglect the competition among individuals and equate the target population size $N$ with the mutant reproduction factor for $T_{\rm fix}$ consecutive generations. Combining Eqs.~(\ref{exp}) and (\ref{m2})
we find $T_{\rm fix} \approx \log N /(\epsilon x)$ in the absence of competing mutations.
The average time between successful mutations is $\lambda_x^{-1}$.
So the condition ensuring 
non-overlapping
sweeps is that the product $\lambda_x T_{\rm fix} \approx \mu N \log N e^{-x^2}$ must be very small so that
\begin{equation}
 \mu N \log N \ll 1\,.
\label{12}
\end{equation}
In other words, 
condition Eq.~\eqref{12} guarantees that after a mutation the new set of types settles in an equilibrium before the next successful mutation event. 

\subsection{Location of the first evolutionary branching}\label{sec:1B}
The trait substitution process considerably slows down as $x$ approaches zero,
so that it must be considered as a sequence of discrete jumps of size $\epsilon$ towards zero, rather
than a continuous process. At levels close to zero a more refined analysis of fitness asymptotics 
is required.

In the following we derive an approximation for the location
of the first evolutionary branching. We show that it happens close to zero, but
typically not at $x=0$. More precisely, we demonstrate that evolutionary
branching becomes possible in an interval of width of order $\epsilon$ around 
$x=0$. It is thus necessary to distinguish between
trait values separated by a few mutation steps. 
For simplicity we assume in the following that initially $x >0$, so that
the trait substitution process approaches $x=0$ from above (as in 
the simulation results shown in Fig. \ref{Fsim}). The problem of
determining the location of the first branching event is symmetric and corresponding expressions for $x < 0 $ are easily found.

The aim is to find trait values $x > 0$ where mutations to and from $x-\epsilon$ are mutually invasive.
The corresponding condition is:
$M_{x-\epsilon}^{\{x\}}>1$ and $M_{x}^{\{x-\epsilon\}}>1$.
Using Eq.~\eqref{1eps} and dropping the terms much smaller than $\epsilon^2$ we arrive at two conditions:
\begin{eqnarray}
M_{x-\epsilon}^{\{x\}}-1&\approx& x\epsilon+\alpha\epsilon^2/2>0\,,\quad\mbox{and}\nonumber\\\ M_{x}^{\{x-\epsilon\}}-1&\approx& 
-(x-\epsilon)\epsilon+\alpha\epsilon^2/2>0\,.
\label{2b}
\end{eqnarray}
We conclude that the first evolutionary branching is possible for
$0 < x < (1+\alpha/2)\epsilon$, so that populations with trait values $x$
and $x-\epsilon$ are mutually invasive provided:
\begin{equation}
\label{eq:1stbrc}
\epsilon \le x \leq j^\ast\epsilon\,\quad\mbox{with}\quad j^\ast = \lceil\alpha/2\rceil\,.
\end{equation}
Here $ \lceil\alpha/2\rceil$ denotes the smallest integer greater than $\alpha/2$. (To avoid complications arising in the situation when 
$j^\ast=1+\alpha/2$  and $M_{x}^{\{x-\epsilon\}}-1=o(\epsilon^2)$ for $x = j^\ast\epsilon$, we will assume that $\alpha$ is not an even integer.) The fact that the critical trait value $j^*\epsilon$ is larger for larger values of $\alpha$ is very intuitive: as competition becomes weaker, initially impossible invasion of $(j^*-1)\epsilon$ into $j^*\epsilon$  becomes favorable once the benefit of reduced competition outweighs the cost of having a "worse" trait value.

\begin{figure}[t]
\includegraphics[width=80mm]{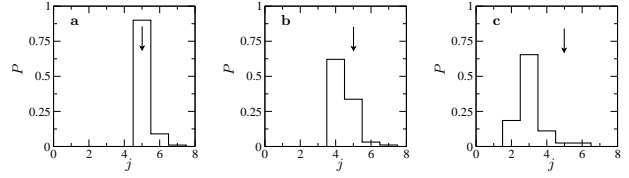}
\caption{Shows histograms of the locations $j$ of the first branching event. All histograms are normalised to unity.
Initial condition $j=7$. Parameters: $\alpha=9$, $\epsilon = 10^{-2}$. The first evolutionary branching
is expected to occur in the interval $-5 \leq j \leq 5$. The upper limit of
this interval is indicated by black arrows.
Panel {\bf a}: $N=5\times 10^6$, $\mu=2\times 10^{-10}$, all 100 runs resulted in an evolutionary branching.
Panel {\bf b}: $N=10^6$, $\mu=10^{-9}$, out of total 100 total only 85 runs resulted in an evolutionary branching while 15 runs  had no branching during the simulation time.
Panel {\bf c}: $N=10^5$, $\mu=10^{-8}$, out of total 100 total only 81 runs succeeded and 19  had no branching during the simulation time. }
\label{fig:hist}
\end{figure}

Fig.~\ref{fig:hist} shows where the first evolutionary branchings occurred in an ensemble
of 300 simulations of the stochastic, individual-based model. The parameter $\alpha$ was
chosen to be $\alpha=9$ which implies that  $j^\ast = 5$. The parameters were chosen so that the
overall substitution rate was the same for all runs, $\theta = \mu N = 10^{-3}$.
All simulations were started at $x_0=7\epsilon$, and were run for the same total time $t$,
given by $\theta t = 10^5$. In a small number of runs evolutionary branching did not occur
during this time. Fig.~\ref{fig:hist} shows where the first branching occurred for the remaining runs.
First, we see that apart from a small number of cases, branching occurs in the region predicted.
Second, the larger the population size $N$ is, the more likely it is that
evolutionary branching occurs at the boundary of the branching region ($j^\ast=5$ in this case).
To explain this behaviour requires a more refined analysis. We return to this
question below in Section \ref{sec:3.5}. 
Third, in the limit $\epsilon \rightarrow 0$ the condition
(\ref{eq:1stbrc})  is consistent with the standard prediction 
in this limit (that the first branching occurs at $x=0$). Fourth, some branchings occurred one mutation step earlier than expected. This can be explained by the fact that occasionally it can take a long time for a slightly advantageous mutant to
stabilise, long enough for the next mutation to initiate the first evolutionary branching.

\subsection{Critical state of coexistence prior to first branching}
\label{sec:3.5}
\begin{figure}[t]
\includegraphics[width=80mm]{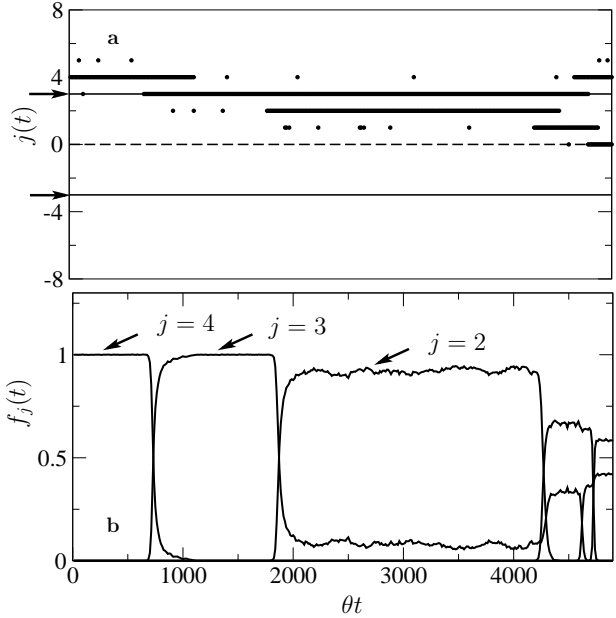}
\caption{Shows coexistence of trait values prior to the first branching. {\bf a} Evolution of the integer trait value $j(t)$
as a function of time. Parameters: $\alpha = 5$, $\epsilon = 10^{-2}$,
$N=5\times 10^6$, $\mu=10^{-9}$. The branching region
is given by $-3 \leq j \leq 3$. {\bf b} Dependence of
$f_j(t)$ upon time. Clearly seen is the coexistence of the trait
values $j=2,3$ at $f_2 = 11/12$ and $f_3 = 1/12$. In the case
shown, coexistence lasts sufficiently long for evolutionary branching
to occur.}
\label{fig:fjt}
\end{figure}

In the preceding section we have derived conditions determining
where the first evolutionary branching may occur,
Eq.~(\ref{eq:1stbrc}).
However, results of our computer simulations of the stochastic model described
in Section \ref{sec:model} also demonstrate that while evolutionary branching
may occur (and often does occur) when these conditions for evolutionary branching are met,
branching need not necessarily take place immediately. This raises the question how long it takes
on average in a finite population for evolutionary branching to occur (a question also
raised by  \citet{B}).
What is the probability that evolutionary branching happens when the substitution
process reaches the boundary of the region where the first evolutionary branching becomes possible
(Fig. \ref{fig:hist})?
Fig. \ref{fig:fjt} shows a realisation of a computer simulation of this situation
for $\alpha=5$ (corresponding to $j^\ast = 3$).
In Fig. \ref{fig:fjt}{\bf a} we see how the trait value reaches the critical value $j^\ast=3$
from above, and how subpopulations with trait values $j=3$ and $j=2$ coexist until
evolutionary branching occurs. The corresponding population sizes $f_j$ are shown
in Fig. \ref{fig:fjt}{\bf b}. For evolutionary branching to occur, the critical state of
coexistence must last sufficiently long for a favourable mutation to occur.
It is possible, however, that the critical state of coexistence may spontaneously
disappear by a fluctuation. We see in Fig. \ref{fig:fjt}{\bf b} that the population-size
fluctuations in the coexistence state are substantially larger than the population-size
fluctuations in the monomorphic states. 

\begin{figure}[t]
\includegraphics[width=60mm]{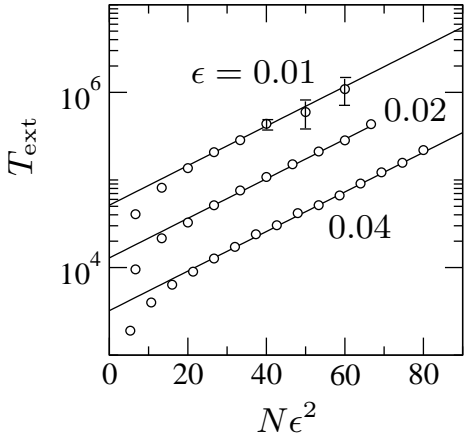}
\caption{\label{fig:text}
Shows time $T_{\rm ext}$ to extinction of the coexistence state
at the boundary of the branching region for the first
branching, in the absence of mutations (symbols).
Parameters: $\mu=0$, $\alpha =3$, $\epsilon = 0.01$, $0.02$, and $0.04$.
Unless otherwise indicated,
two-sigma error bars are smaller than the symbol size.
The error bars were determined from 1000 independent simulations,
with three exceptions: for $\epsilon = 0.01$ and $N\epsilon^2 = 40$,
100 independent simulations were used, for $N\epsilon^2=50$ and $60$,
20 independent simulations.
Also shown are fits to the law (\ref{eq:text}), lines.
We find $S_3 \approx 0.052$.}
\end{figure}

A preliminary analysis shows that
while the fluctuations of the monomorphic population sizes are of order $N^{-1/2}$,
the fluctuations of the subpopulation sizes in the critical state of coexistence are much larger: of order $(N \epsilon^2)^{-1/2}$
in the limit of large values of $N$ and small values of $\epsilon$.
The smaller value $\epsilon$ takes, the shorter is the average life time
of the critical state of coexistence, lowering the probability that evolutionary
branching occurs in this state. Without going into details, we just briefly sketch our argument (which remains to be made rigorous). Let  $(f_k,g_k)$ be the densities $(f_{j\epsilon}^{\{j\epsilon,j'\epsilon\}},f_{j'\epsilon}^{\{j\epsilon,j'\epsilon\}})$ of two populations coexisting at time $k$ and having the neighbouring trait values $j\epsilon$ and $j'\epsilon$ with $j'=j-1$. The evolution of the vector $(f_k,g_k)$ can be approximated by a stochastic dynamical system
\begin{align}\label{sd}
\left\{ 
\begin{array}{l}
f_{k+1}=\frac{2f_k}{1+f_k+ag_k}+{1\over\sqrt{N}}u_k,\quad  a=e^{-\epsilon^2(\alpha+2-2j)},\\
g_{k+1}=\frac{2g_k}{1+bf_k+g_k}+{1\over\sqrt{N}}w_k,\quad  b=e^{-\epsilon^2(\alpha+2j)},
\end{array}
\right.
\end{align} 
where $u_k$ and $w_k$ are independent normal random variables with zero means and standard deviations ${ 2\sqrt{f_k(f_k+ag_k)} \over1+f_k+ag_k} $ and $ { 2\sqrt{g_k(bf_k+g_k)} \over1+bf_k+g_k} $ respectively. Provided the deterministic part has a nontrivial stable point we consider a linearised version of this system around this stable point. Our analysis indicates that while the sum $f_k+g_k$ behaves as a stationary autoregression process with fluctuations of the order $N^{-1/2}$, the difference $f_k-g_k$ behaves as a stationary autoregression process with fluctuations of order  $(N \epsilon^2)^{-1/2}$.
Given that the fluctuations in the sizes of coexisting populations featuring the trait values $(j^*\epsilon,(j^*-1)\epsilon)$ are of order   $(N \epsilon^2)^{-1/2}$ a rough estimate of the probability of the sudden loss of coexistence is obtained using the approximate tail probability for the normal distribution $c_1e^{-c_2N \epsilon^2}$, where $c_1$ and  $c_2$ are positive constants.
Thus, the life time $T$ of the critical state of coexistence is expected to be inversely proportional to the probability of sudden loss of coexistence $c_1e^{-c_2N \epsilon^2}$ so that
 \begin{equation}
\label{eq:text}
\log T \sim S_\alpha N \epsilon^2\,
\end{equation}
for some constant $S_\alpha$ depending on $\alpha$ through \eqref{eq:1stbrc}.  
Fig.~\ref{fig:text} shows simulation results for the life time $T$ as a function of $N$ and $\epsilon$.
The results of the simulations are consistent with the expectation Eq.~(\ref{eq:text}).
We find $S_3 \approx 0.052$.
The precise mathematical derivation of Eq.~(\ref{eq:text}) and the
calculation of the constant $S_\alpha$ are interesting questions 
for further work.

\subsection{Evolution after the first branching}
\label{sec:2B}
Condition Eq.~(\ref{eq:1stbrc}) demonstrates that if
the first evolutionary branching occurs, then it happens within 
a region of size $\epsilon$ around $x=0$.
We now turn to the evolution of the pair of trait values 
$(x_1,x_2)$ after the first evolutionary branching.
Standard theory suggests that the dimorphic population
evolves symmetrically (along the dashed line in Fig. \ref{Fsim2}).
Our simulations of the stochastic, individual-based population model with discrete mutational steps
show that the pair of coexisting trait values followed through the chain of consecutive replacements does not develop in a fully symmetrical fashion. 
The random path on the $x_1$-$x_2$-plane  (see Fig.~\ref{Fsim2}) follows the trajectory of a random walk with steps from $(x_1,x_2)$ to either $(x_1,x_2+\epsilon)$ or $(x_1-\epsilon,x_2)$ depending on which of the two branches the next replacement has been successful. The repulsion between the two branches is due to the pressure of  competition. It is advantageous to stay further apart reducing interspecies competition.

Observe that the consecutive steps in this random walk are weakly negatively dependent: the further the walk deviates from the diagonal $x_1=-x_2$, the larger is the probability for the next step to decrease this deviation. To see this, consider a pair of coexisting populations having trait values  $x_1$ and $x_2$ with $x_1<0< x_2$ and for definiteness  $x_2>|x_1|$. 
For a given branch, the corresponding replacement rate  is a product of two factors (see the discussion leading to Eq.~\eqref{eq:lambdax} in the monomorphic case): the stable population size and the probability of fixation for a single mutant. We show that given the lower branch is closer to zero, $x_2>|x_1|$, both factors of the replacement rate for the lower branch are larger making the move of the lower branch (always off zero, see Appendix B) more probable. Indeed, as the negative trait value is closer 
to zero, its stable population size must be larger than the size of the population with the positive trait value. On the other hand, as computed in Appendix B, the probability of fixation for a single mutant in the $x_1$-population is also larger, Eq.~\eqref{su3}.

The last observation implies that with high probability the locations of the pair of branches satisfy
\be
x_1=-x_2+O(\sqrt{\epsilon})
\label{O}
\ee
prior to the second evolutionary branching. This means that the the expected location of the second evolutionary branching, if any, should not deviate from the diagonal more than a constant times $\sqrt{\epsilon}$.
Indeed, the predicted deviation from the diagonal can only be made larger if we neglect the negative dependence and consider a symmetric random walk with independent steps. Referring to the classical central limit theorem for the simple symmetric random walk we observe that the quantity
$x_1+x_2$ (which is always zero in the symmetric case) is of order $\epsilon\sqrt{n}$,
where $n$ is the number of jumps in the random walk (representing the number of replacements since the first evolutionary branching).
Since the second evolutionary branching takes place
at values of $x_1$ and $x_2$ of order unity, one must wait for a large number of jumps,
namely of order $1/\epsilon$ for the second branching to occur.
We conclude that the path taken by the two branches
in the $x_1$-$x_2$ plane will deviate from the diagonal $x_1=-x_2$ 
by a distance of at most of order $\epsilon\sqrt{1/\epsilon}=\sqrt{\epsilon}$.  

\subsection{\lq Triggering cross \rq for the second branching}
\label{sec:2B1}

We now discuss the location of the second evolutionary branching.
As in the previous section, this location is determined by mutual invasibility
analysis, but now for the dimorphic case $D=\{x_1,x_2\}$. 
The second evolutionary branching may occur on either of the two branches stemming
from the first evolutionary branching. We first discuss
the condition for the case where the second evolutionary branching occurs on the upper branch 
of the dimorphic population (an example is shown in Fig.~\ref{Fsim}{\bf a}).
With $z=x_2\pm\epsilon$ the corresponding conditions for mutual invasibility are $M_{z}^{\{x_{1},x_{2}\}}>1$ and $M_{x_{2}}^{\{x_{1},z\}}>1$.
Analysis of the signs of $M_{z}^{\{x_{1},x_{2}\}}-1$ and $M_{x_{2}}^{\{x_{1},z\}}-1$ , reported in Appendix B, shows that in view of Eq.~\eqref{O} the 
second evolutionary branching of the upper branch becomes possible in the region
\be
x_1=-x_\alpha+O(\sqrt{\epsilon}), \; x_2=x_\alpha+O(\sqrt{\epsilon}),
\label{O1}
\ee
where $x_\alpha$ is given by Eq.~\eqref{xa}.
More precisely, we show (see Appendix C) that the second evolutionary branching of the upper branch with high probability occurs in the region
\begin{equation}\label{au1}
(x_1+x_\alpha) c_\alpha- x_2+x_\alpha=O(\epsilon)
\end{equation}
which is a neighbourhood of the straight line 
\begin{equation}
\label{e1}
(x_1+x_\alpha)c_\alpha=x_2-x_\alpha.
\end{equation}
The coefficient  $c_\alpha$ is given by
\begin{equation}
c_\alpha={(1+2\alpha)^2\log(1+2\alpha)-2\alpha(1+\alpha)\over (1+2\alpha)(2\alpha-1)\log(1+2\alpha)+2\alpha(1+\alpha)}\label{ca}\,.
\end{equation}
It is shown in Appendix A that $c_\alpha$ satisfies $c_\alpha < 1$. The minimum value of $c_\alpha$ equals $0.7732\dots$ and is achieved at $\alpha=4.0533\dots$.

Turning to the possibility that the second branching occurs
on the lower branch we find a similar condition 
\begin{equation}\label{au2}
(x_2-x_\alpha) c_\alpha- x_1-x_\alpha=O(\epsilon)
\end{equation}
corresponding to an $\epsilon$-neighbourhood of another straight line 
\begin{align}
&x_1+x_\alpha=(x_2-x_\alpha)c_\alpha\label{e2}\,.
\end{align}

\begin{figure}[t]
\includegraphics[width=60mm]{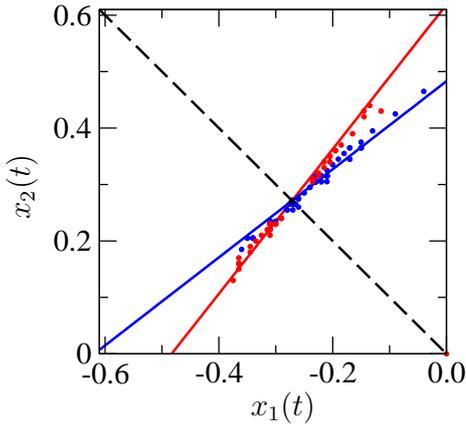}
\caption{Locations of the second evolutionary branching in the $x_1$-$x_2$-plane.
Shown are the lines (\ref{e1}), solid blue, and (\ref{e2}), solid red, as well as the line corresponding to symmetric evolution of $(x_1,x_2)$ (black dashed).
Results of simulations determining the locations of second branchings
are shown as points.  Blue points correspond to cases where
the second branching occurred on the upper branch, red points
to branchings of the lower branch.
Parameters: $\alpha=9$, $\epsilon = 10^{-2}$,
$N=4\times 10^6$, $\mu=2.5\times 10^{-10}$, $\theta t \leq 4 \times 10^3$.
Six sets of initial conditions were used: $(j_1,j_2) = (-20,30), (-30,20), (-10,35), (-25,25), (-35,10)$, and $(-1,45)$. }
\label{fig:pd2}
\end{figure}

Taking these results together, we have shown that
the second evolutionary branching is expected to occur in the vicinity
of the lines  Eqs.~\eqref{e1} and  \eqref{e2}.
If the pair  $(x_1,x_2)$ 
of diverging branches comes close to the line Eq.~\eqref{e1},
then a second evolutionary branching of the upper branch becomes possible.
Conversely, when the pair $(x_1,x_2)$ comes close 
to the line Eq.~\eqref{e2}, a second evolutionary branching may occur on the lower
branch. 
Note that since $\mu$ is assumed to be independent of the trait value, our answers Eqs.~\eqref{e1} and  \eqref{e2} depend only on $\alpha$.
The cross formed by the pair of lines \eqref{e1}, and \eqref{e2} is centered 
at the point $(-x_\alpha,x_\alpha)$ in the $x_1$-$x_2$-plane. 

Eqs.~\eqref{e1}, \eqref{e2} are consistent with results of direct numerical simulations
of the model. Fig.~\ref{fig:pd2} shows the lines (\ref{e1}) and (\ref{e2}) for $\alpha=9$ 
and $\epsilon=10^{-2}$, as well as the locations of second evolutionary branchings 
for an ensemble of simulations of the stochastic, individual-based model. 
Six sets of dimorphic initial conditions were used, $(x_1,x_2) = 
(-0.2,0.3), (-0.3,0.2), (-0.25,0.25), (-0.35,0.1), (-0.1,0.35)$ and $(0.01,0.45)$.
We remark that the last three initial conditions deviate substantially from
the diagonal (dashed line in Fig.~\ref{Fsim2}). As shown above, paths in
the $x_1$-$x_2$-plane typically exhibit deviations of order $\sqrt{\epsilon}$ 
from the diagonal. We have nevertheless included dimorphic initial
conditions far from the diagonal in order to clearly exhibit the locations
of the second evolutionary branching.

In Fig.~\ref{fig:pd2}, blue dots correspond to cases where the second evolutionary
branching occurred on the upper branch of the dimorphic state, while red
dots correspond to cases where the second evolutionary branching
occurred on the lower branch. The coordinates of the dots show the location
of the second branching event in the $x_1$-$x_2$-plane (we determined
this location by taking the average of the trait values of the critical
state of coexistence prior to the second branching, as well as the 
coordinate of the second branch when coexistence first occurred).
Fig.~\ref{fig:pd2} is consistent with the 
theoretical expectation
that second branchings of the upper branch 
are triggered by the line  Eq.~\eqref{e1}, while the 
second branchings of the lower branch are triggered by the line  Eq.~\eqref{e2}.

\section{Conclusions}
\label{sec:conc}
We have analysed evolutionary branching in a stochastic, individual-based
model for the evolution of a trait value subject to small but non-negligible mutational step sizes $\epsilon$.
We found that the branching patterns are in general asymmetric at
distinct, non-null
mutational steps, complementing 
the standard adaptive dynamics predictions \cite{G98}
valid in the limit $\epsilon \rightarrow 0$. In particular we found
conditions describing
the locations where the first branching may occur, Eq.~(\ref{eq:1stbrc}), and where
the second branching may take place, Eqs.~\eqref{O1}, \eqref{au1}, and \eqref{au2}. Results of simulations of
a stochastic, individual-based model at small mutational step sizes $\epsilon$ 
were seen to be consistent with these conditions.
Our results are derived in the limit of large population sizes $N$ and small
mutation rates $\mu$, but allow for fixed mutational step sizes.  In
this respect our results 
Eqs.~(\ref{eq:1stbrc}), \eqref{O1}, \eqref{au1}, and \eqref{au2}
 complement the established approach. 

We have also demonstrated that population-size fluctuations in the critical
state of coexistence prior to a branching may erase this
state. Indeed, its sojourn time depends sensitively
on  mutational step size. Evolutionary branching occurs typically
only when this time is much longer than the time between successful mutations.

\section*{Acknowledgments}
Financial support from the Swedish Research Council (SS, BM), the G\"oran Gustafsson Foundation
for Research in Natural Sciences and Medicine (BM), by the Bank of Sweden Tercentenary Foundation (SS) is
gratefully acknowledged. 
VV was partially supported by the grant RFBR 11-01-00139 
and the program "Dynamical systems and control theory" of the Russian Academy of Sciences.

\section*{Appendix A}
This appendix is divided into three sections summarising material needed in appendices B and C.

{\bf 1}. We discuss the properties of the three functions:
\begin{align}
&C(\alpha)={\alpha\over2}-((1+\alpha)^2-\alpha^2)x_\alpha^2,\label{Ca}\\
&c_1(\alpha)={(1+2\alpha)^2\over2\alpha(1+\alpha)}\log(1+2\alpha)-1,\label{c1a}\\
&c_2(\alpha)={(1+2\alpha)(2\alpha-1)\over2\alpha(1+\alpha)}\log(1+2\alpha)+1\,.\label{c2a}
\end{align} 
Here $x_\alpha$ is defined by Eq.~\eqref{xa}. 
We show that $C(\alpha)>0$, and $0<c_1(\alpha)<c_2(\alpha)$ for all positive 
values of $\alpha$. The latter inequalities in view of $c_\alpha=c_1(\alpha)/c_2(\alpha)$ imply that the constant $c_\alpha$, defined by Eq.~\eqref{ca} and determining the slopes of the triggering lines Eqs.~\eqref{e1}, \eqref{e2}, satisfies $0 < c_\alpha < 1$.

To see that $C(\alpha)>0$ we observe that
\begin{align}
C(\alpha)&={\alpha\over2}-{1+2\alpha\over4(1+\alpha)}\log(1+2\alpha)=\frac{h(1+2\alpha)}{8(1+\alpha)}\,,
    \end{align}
where $h(x)=x^{2}-2x\log x-1$. The function $h(x)$ takes positive values for all $x>1$, since $h(1)=0$ and $h^{\prime }(x)>0$ for $x>1$. Thus $C(\alpha)>0$ 
for all positive $\alpha$.

Similarly, the fact that $0<c_1(\alpha)<c_2(\alpha)$ follows from
\begin{align}
c_1(\alpha)={h_1(1+2\alpha)\over2\alpha(1+\alpha)}\,,\quad c_2(\alpha)=c_1(\alpha)+{4\over\alpha}C(\alpha)\,,
\end{align}
where $h_1(x)=x^2\log x-\frac{x^{2}-1}{2}$ is positive for $x>1$ (since $h_1(1)=0$ and $h'_1(x)>0$).

{\bf 2}. We discuss  properties of the equilibrium densities 
in the dimorphic case $D=\{x_1,x_2\}$. In this case, 
the equilibrium densities $f_{x_2}^{\{x_1,x_2\}}$ and $f_{x_1}^{\{x_1,x_2\}}$ satisfying the system Eq.~(\ref{MD1}):
\begin{align*}
&R_{x_1x_2}f_{x_2}^{\{x_1,x_2\}}+f_{x_1}^{\{x_1,x_2\}}=1,\\
&f_{x_2}^{\{x_1,x_2\}}+R_{x_2x_1}f_{x_1}^{\{x_1,x_2\}}=1,
\end{align*}
can be represented as $f_{x_2}^{\{x_1,x_2\}}=\phi(x_1,x_2)$ and $f_{x_1}^{\{x_1,x_2\}}=\phi(x_2,x_1)$ with
\begin{align}
\phi(x_1,x_2)={1-R_{x_2x_1}\over 1-R_{x_1x_2}R_{x_2x_1}}={1-e^{-(x_2-x_1)(\alpha x_2-(2+\alpha)x_1)}\over1-e^{-2(1+\alpha)(x_1-x_2)^2}}.
\label{fi}
\end{align}
In particular, in the symmetric case $x_1=-x$ and $x_2=x$, we obtain 
\begin{align}\label{sym}
   \phi(-x,x)=\phi(x,-x)={1\over1+e^{-4(1+\alpha)x^2}}.
\end{align}

{\bf 3}. We quote the Taylor expansion for the function defined by Eq.~\eqref{MD} (recall that $M_x^D=1$ for $x\in D$)
\begin{eqnarray}
\label{den2}
M_{z}^D-1&\approx&\left(\alpha x-(1+\alpha)A_{x}^{D}\right)\Delta\\
&+&\left({\alpha\over2} -\alpha^2 x^2+2\alpha(1+\alpha)xA_{x}^{D}-(1+\alpha)^2B_{x}^{D}\right)\Delta^2\nonumber
\end{eqnarray}
valid for small values of $\Delta=z-x$. Here the terms
\begin{align*}
    A_{x}^{D}=\sum_{y\in D}yR_{xy}f_{y}^D,\ \
    B_{x}^{D}=\sum_{y\in D}y^2R_{xy}f_{y}^D
\end{align*}
assume the form of 
first and second moments of the coexisting trait values using the weights $R_{xy}f^D_{y}$  satisfying  
condition Eq.~\eqref{MD1}.

\section*{Appendix B} 
In this appendix we discuss in which region of the $x_1$-$x_2$-plane the second evolutionary branching becomes possible.
It may occur on either of the two branches stemming
from the first evolutionary branching. We start by discussing
the evolution of the upper branch of the dimorphic population on its way towards the second branching (an example is shown in Fig.~\ref{Fsim}{\bf a}).
With $z=x_2\pm\epsilon$ the corresponding conditions for mutual invasibility are $M_{z}^{\{x_{1},x_{2}\}}>1$ and $M_{x_{2}}^{\{x_{1},z\}}>1$.
The signs of $M_{z}^{\{x_{1},x_{2}\}}-1$ and $M_{x_{2}}^{\{x_{1},z\}}-1$ are determined by Eq.~{\eqref{den2} with
\begin{eqnarray}
A_{x_2}^{\{x_1,x_2\}}&=&x_2\phi(x_1,x_2)+x_1R_{x_2x_1}\phi(x_2,x_1)\nonumber\\
&=&x_1+(x_2-x_1)\phi(x_1,x_2)\,,\label{a12}\\
B_{x_2}^{\{x_1,x_2\}}&=&x_2^2\phi(x_1,x_2)+x_1^2R_{x_2x_1}\phi(x_2,x_1)
\nonumber\\
&=&x_1^2+(x_2^2-x_1^2)\phi(x_1,x_2)\,,\label{b12}
\end{eqnarray}
where $\phi(x_1,x_2)$ is given by Eq.~(\ref{fi}).
Using Eqs.~{\eqref{den2}, \eqref{a12}, and \eqref{b12} we find
\begin{align}
M_{z}^{\{x_1,x_2\}}-1\approx C_1(x_1,x_2)(z-x_2)+C_2(x_1,x_2)\epsilon^2\,,
\label{M2}
\end{align}
where
\begin{eqnarray}
C_1(x_1,x_2)&=&\alpha x_2-(1+\alpha)x_1\\\nonumber &-&(1+\alpha)(x_2-x_1)\phi(x_1,x_2)\,,\label{C1}\\
C_2(x_1,x_2)&=&{\alpha\over2}\!-\!\{\alpha x_2-(1+\alpha)x_1\}^2\\\nonumber
&+&(1+\alpha) (x_2\!-\!x_1)\{(\alpha\!-\!1)x_2 \\\nonumber &-&(1+\alpha)x_1\}\phi(x_1,x_2)\nonumber\,.
\end{eqnarray}
Using a counterpart of Eq.~\eqref{spr} for the survival probability $P_{x_2+\epsilon}(x_1,x_2)$ of a single $(x_2+\epsilon)$-mutant in a stable $(x_1,x_2)$-dimorphic population we find that
\be\label{su1}
P_{x_2+\epsilon}(x_1,x_2)\approx2\epsilon (x_2-x_1)(\alpha- (x_2-x_1)^{-1}x_1-(1+\alpha)\phi(x_1,x_2)).
\ee

Similarly, consider the possibility that the second branching occurs
on the lower branch. In this case we have for $z=x_1\pm\epsilon$
\begin{align}\label{muza}
M_{z}^{\{x_1,x_2\}}-1\approx C_1(x_2,x_1)(z-x_1)+C_2(x_2,x_1)\epsilon^2\,
\end{align}
which yields
\begin{eqnarray}
\label{su2}
P_{x_1-\epsilon}(x_1,x_2)&\approx&2\epsilon (x_2-x_1)(\alpha+ (x_2-x_1)^{-1}x_2\\
&&\nonumber -(1+\alpha)\phi(x_2,x_1)).
\end{eqnarray}
Close inspection of Eqs.~\eqref{su1} and \eqref{su2} for $x_1<0<x_2$ using Eq.~\eqref{fi} reveals that
\be\label{su3}
P_{x_1-\epsilon}(x_1,x_2)>P_{x_2+\epsilon}(x_1,x_2) \mbox{ given } |x_1|<x_2.
\ee
As explained in Section \ref{sec:2B}, this is one of the factors ensuring the negative dependence among the steps of the random walk in the $x_1$-$x_2$-plane leading to the  condition Eq.~\eqref{O}.

Applying Eq.~\eqref{O} we can further refine our previous analysis and deduce from Eq.~\eqref{M2}
\begin{equation}
M_{z}^{\{x_1,x_2\}}-1\approx C_1(x_1,x_2)(z-x_2)+C_2(-x_2,x_2)\epsilon^2\,.
\label{M3}
\end{equation}
Replacing $x_1$ by $-x_2$ in this expression incurs  error of order $\sqrt{\epsilon}$. 
This implies:
\begin{equation}
M_{z}^{\{x_1,x_2\}}-1=C_1(-x_2,x_2)(z-x_2)+O(\epsilon^{3/2})\,.
\label{M4}
\end{equation}
It follows from Eq.~\eqref{sym} that
\begin{align}
C_1(-x,x)&=(1+2\alpha)x-{2(1+\alpha)x\over1+e^{-4(1+\alpha)x^2}}\,,\label{C1x}\\
C_2(-x,x)&={\alpha\over2} -(1+2\alpha)^2 x^2+{4\alpha(1+\alpha) x^2\over1+e^{-4(1+\alpha)x^2}}\label{C2x}.
\end{align}
Now let $x_\alpha$ be given by Eq.~\eqref{xa}.
Since $C_1(-x,x)$ is positive for $x\in(0,x_\alpha)$,}
Eq.~\eqref{M4} confirms that the replacement process on the upper branch goes upward until $x_2=x_\alpha+O(\sqrt{\epsilon})$. 

Similarly, consider the possibility that the second branching occurs
on the lower branch. In this case using Eqs.~\eqref{O} and \eqref{muza} we get a lower branch counterpart of Eq.~\eqref{M4}
\begin{equation}
M_{z}^{\{x_1,x_2\}}-1=C_1(x_1,-x_1)(x_1-z)+O(\epsilon^{3/2})\,
\label{M5}
\end{equation}
implying that the replacement on the lower branch goes downward until $x_1=-x_\alpha+O(\sqrt{\epsilon})$. 
We conclude that the 
second evolutionary branching on one of the two branches is only possible in the region Eq.~\eqref{O1}.

\section*{Appendix C}
In this appendix we establish  Eq.~\eqref{au1} in Section \ref{sec:2B1} which in turn implies Eq.~\eqref{e1} determining the triggering line for the upper branch. Given Eq.~\eqref{O1}, the deviations
$\delta_1=x_1+x_\alpha$ and $\delta_2=x_2-x_\alpha$ satisfy $\delta_i=O(\sqrt{\epsilon})$.
We demonstrate first that
\begin{eqnarray}
  C_1(x_1,x_2)&=&{1\over2}c_1(\alpha)\,\delta_1-{1\over2}c_2(\alpha)\,\delta_2+c_{11}(\alpha)\,\delta_1^2\nonumber\\
&+&c_{12}(\alpha)\,\delta_1\delta_2\\\nonumber &+&c_{22}(\alpha)\,\delta_2^2+o(\epsilon)\,,\label{AA}
\end{eqnarray}
where $c_i(\alpha)$ are given by \eqref{c1a} and  \eqref{c2a}, while the exact form of the constants $c_{ij}(\alpha)$ does not matter.
Using
$\phi(-x_\alpha,x_\alpha)={1+2\alpha\over 2(1+\alpha)}$
and the Taylor expansion of Eq.~\eqref{fi} we obtain
\begin{eqnarray}
 \phi(x_1,x_2)&=&{1+2\alpha\over 2(1+\alpha)}+\delta_1\phi_1+\delta_2\phi_2+{\delta_1^2\over2}\phi_{11}\nonumber\\
&&+{\delta_1\delta_2\over2}\phi_{12}+{\delta_2^2\over2}\phi_{22}+o(\epsilon)\,,
\end{eqnarray} 
where $\phi_{i}={\partial\over\partial x_i}\phi(x_1,x_2)\vert_{x_1=-x_\alpha,x_2=x_\alpha}$ and $\phi_{ij}$ are the corresponding second order derivatives. It follows in view of Eq.~\eqref{C1} ,
that Eq.~\eqref{AA} holds with
\begin{eqnarray}
 c_1(\alpha)&=& -1-4(1+\alpha)x_\alpha\phi_1,\label{c1a1}\nonumber\\
 c_2(\alpha)&=& 1+4(1+\alpha)x_\alpha\phi_2,\label{c2a1}\nonumber\\
 c_{11}(\alpha)&=& (1+\alpha)(\phi_1-x_\alpha\phi_{11}),\\
 c_{12}(\alpha)&=& (1+\alpha)(\phi_2-\phi_1- x_\alpha\phi_{12}),\nonumber\\
c_{22}(\alpha)&=& -(1+\alpha)(\phi_2+ x_\alpha\phi_{22}),\nonumber
\end{eqnarray}
so that it remains only to verify that the new relations for $c_i(\alpha)$ agree with Eqs.~\eqref{c1a} and \eqref{c2a}. 
For this it is sufficient to note that $\phi_1=-{(1+2\alpha)^2x_\alpha\over 2\alpha(1+\alpha)}$ and
$\phi_2={(1+2\alpha)(2\alpha-1)x_\alpha\over 2\alpha(1+\alpha)}$. This follows from Eq.~\eqref{fi}. We conclude that, given Eq.~\eqref{O1}, we have 
\begin{eqnarray}
  2C_1(x_1,x_2)&=&c_1(\alpha)(x_1+x_\alpha)\nonumber\\
&&-c_2(\alpha)(x_2-x_\alpha)+O(\epsilon)\,.\label{AA1}
\end{eqnarray}

To conclude our derivation of Eq.~\eqref{au1} we notice that in accordance with Eqs.~(\ref{M3}) and \eqref{O1} we have
\begin{equation}
M_{z}^{\{x_1,x_2\}}-1\approx C_1(x_1,x_2)(z-x_2)+C(\alpha)\epsilon^2\,,
\end{equation}
where $C(\alpha)=C_2(-x_\alpha,x_\alpha)$ is a positive constant given by Eq.~\eqref{Ca}. Eq.~\eqref{M4} implies that  the inequality $M_{x_2+\epsilon}^{\{x_1,x_2\}}>1$ is equivalent to
$C_1(x_1,x_2) >-\epsilon C(\alpha)$.
Similarly, due to Eq.~\eqref{M4}, the inequality $M_{x_2}^{\{x_1,x_2+\epsilon\}}>1$ translates into
$C_1(x_1,x_2) <\epsilon C(\alpha)$. Combining the last two relations with  Eq.~(\ref{AA1}) we derive Eq.~\eqref{au1}. In other words, second
evolutionary branching on the upper branch of the dimorphic population 
occurs in an $\epsilon$-neighbourhood of the line given by Eq.~(29).

Using  Eq.~\eqref{muza} and repeating the arguments summarised above, 
we find the condition (31) for the second evolutionary branching 
to occur on the lower branch of the dimorphic population.
This branching occurs in an 
$\epsilon$-neighbourhood of the straight line 
Eq.~(32).

\end{document}